# High-pressure synthesis of superhard and ultrahard materials

Yann Le Godec,[1,*]  Alexandre Courac [1]  and  Vladimir L. Solozhenko [2]

[1]  *Institut de Minéralogie, de Physique des Matériaux et de Cosmochimie (IMPMC), Sorbonne Université, UMR CNRS 7590, Muséum d'Histoire Naturelle, IRD UMR 206, 75005 Paris, France*

[2]  *LSPM–CNRS, Université Paris Nord, 93430 Villetaneuse, France*

**Abstract** – A brief overview on high-pressure synthesis of superhard and ultrahard materials is presented in this tutorial paper. Modern high-pressure chemistry represents a vast exciting area of research which can lead to new industrially important materials with exceptional mechanical properties. This field is only just beginning to realize its huge potential, and the image of "terra incognita" is not misused. We focus on three facets of this expanding research field by detailing: (i) the most promising chemical systems to explore (i.e. "where to search"); (ii) the various methodological strategies for exploring these systems (i.e. "how to explore"); (iii) the technological and conceptual tools to study the latter (i.e. "the research tools"). These three aspects that are crucial in this research are illustrated by examples of the recent results on high pressure – high temperature synthesis of novel super- and ultrahard phases (orthorhombic $\gamma$-$B_{28}$, diamond-like $BC_5$, rhombohedral $B_{13}N_2$ and cubic ternary B–C–N phases). Finally, some perspectives of this research area are briefly reviewed.

## I. Introduction

Hardness is one of fundamental characteristics of material and can be described as resistance to various kinds of deformation that material can undergo under mechanical tests or in real life. For quantitative characterization in modern material's science, we use the diamond-indenter penetration test, thus formally the hardness may be referred to the resistance of the surface to mechanical deformation. The ratio of the penetration force to the area of indentation mark on the surface becomes constant at high enough applied forces (the particular importance of this issue for

---
* Corresponding author: yann.le_godec@sorbonne-universite.fr

superhard- and ultrahard materials is crucial [1]), has a dimensionality of pressure, and defined as a hardness named after the type of diamond pyramid used for indentation (so-called Vickers, Knoop or Berkovich hardness, typically used for superhard materials).

Superhard and ultrahard materials can be defined as having Vickers microhardness ($H_V$) exceeding 40 GPa and 80 GPa respectively [2, 3]. In addition to high hardness, they usually possess other unique properties such as compressional strength, shear resistance, large bulk moduli, high melting temperatures and chemical inertness. This combination of properties makes these materials highly desirable for a number of industrial applications. Historically, the first high-pressure experiments designed to produce materials for industrial use were carried out during the second half of the 20th century with the laboratory synthesis of superhard materials, namely, diamond [4, 5] and cubic boron nitride ($c$-BN or $zb$-BN to denote its zinc-blende structure) [6, 7]. Nowadays, the chemical industries linked to these materials are flourishing all over the world with an annual production of 3 000 million carats (1 carat = 0.2 g). Industrial applications of bulk superhard materials to date have been dominated by superabrasives, such as stone and concrete sawing, cutting and grinding tools, polishing tools, petroleum exploration mining, high speed machining of various engineering materials, etc. Recent achievements in search for novel superhard materials indicate that synthesis of phases – other than carbon allotropes, which are of primary interest to this manual – with hardness exceeding that of various forms of diamond (Knoop hardness 56-115 GPa for different *hkl* index planes of natural single-crystals [8] and 120-145 GPa for nanocrystalline diamond [9]) is very unlikely [10]. At the same time, the hardness and mechanical properties of diamond-based materials themselves can still be improved by microstructure control (dislocation density or nanotwinning engineering [11-13]) or by high temperature and pressure treatment (e.g. annealing of Chemical vapor deposition (CVD) diamonds) [14]. Rather than harder, one should consider the possibility to synthesize materials that are more useful, i.e. thermally and chemically more stable than diamond, and harder than cubic boron nitride. Actually, the superabrasive performance of diamond is somewhat limited: it is not stable in the presence of oxygen at even moderate temperature, so diamond cannot be used for very high speed machining; and it is not a suitable abrasive for cutting and polishing ferrous components, because it tends to form iron carbides. In these cases, $c$-BN is a good substitute, but its hardness is only 50 % that of diamond. So, the main motivation of the high pressure exploration of new super- and ultrahard materials remains the search for new materials, which could be more thermally and chemically stable than diamond, but remarkably harder than $c$-BN.

In the present tutorial, since it is not intended as a comprehensive and exhaustive review which has been already documented in many useful articles [15, 16] or book [17], we aim to focus on three aspects of this expanding research field. We try to rationalize and explain in a pedagogical way the methodological keys used in the various studies of the field. Hence, we give references whenever relevant, but limit them to selected ones which are most directly connected to the illustration of our purpose. Also, we prefer illustrate the subjects using concrete examples from our own studies (both published and unpublished) on high pressure – high temperature synthesis of novel superhard phases in the B–C–N–O system (Fig. 1).

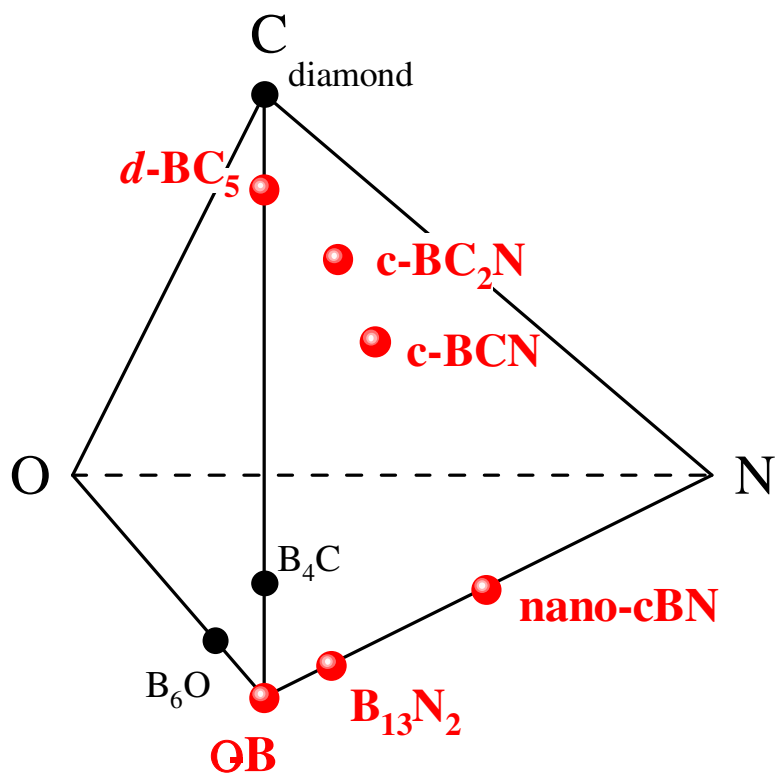

Fig. 1  Superhard phases of the B–C–N–O quaternary system. Phases discovered in last two decades and discussed in the present paper are shown in red.

This tutorial paper is organized as follows. Section II gives an explanation and description of the most promising chemical systems to explore for finding new superhard (including ultrahard) materials. In these systems, three methodological exploring strategies at high pressure are discussed and illustrated by our recent studies in section III. Section IV presents the various experimental and theoretical tools in this field so that the non-specialist readers can get a general idea of the studies on the subject. Some perspectives are given in section V.

## II. Most promising chemical systems to find new superhard and ultrahard materials.

High hardness is determined by hard 3-dimensional networks of strong (and short) covalent bonds, like in the case of period II elements. The rigidity of bonds itself is important characteristics. At the same time, if hard covalent bonds do not form 3-D framework, high hardness can be hardly attainable. An illustration of this is graphite ($H_V$ ~ 0.1 GPa or ~1 GPa depending on crystallographic direction [18]) that has a layered structure (along crystallographic $c$-direction) with weak van-der-Waals bonds, which render it quite soft material, despite the fact that in-layer bonds are stronger – in the terms of energy – than that in diamond. The detailed impact of these factors has been recently described in a review on the theoretical design of superhard and ultrahard materials [19]. From the point of view of experimental material designer, the most important parameters are the composition, crystal structure, together with the ($p$, $T$) conditions that can be explored. That is why here we will principally consider the compositions and structural types already known for their super- and ultrahardness.

So far, boron and carbon are the hardest elements known, and therefore it comes naturally to search for new super- and ultrahard phases among their compounds [20]. Carbon allotropes has been widely explored by *ab initio* search algorithms [21, 22], and some of them have been observed experimentally like so-called monoclinic M-carbon [23]. A number of boron allotropes has been also predicted. In this tutorial we will describe the synthesis of one high pressure allotrope of boron as an example.

Borides are another large family of promising superhard compounds [2]. For example, boron-rich phases of the B–C–N–O(–X) system with an α-rhombohedral-boron (α-$B_{12}$) like structure (boron carbide $B_4C$, boron suboxide $B_6O$, etc.) combine high hardness and wear resistance, chemical inertness, high melting point as well as high cross-section for neutron absorption [20]. Recently β-form of boron suboxide, orthorhombic β-$B_6O$ with *Cmcm* space group has been predicted [24], with higher hardness than the known hard and low-compressible rhombohedral α-$B_6O$ [25].A number of new superhard B-C-O compounds of known and hypothetical structural types has been also predicted recently, tetragonal thermodynamically stable and potentially superhard phase $B_4CO_4$, and two low-enthalpy metastable compounds (like $B_6C_2O_5$, $B_2CO_2$) [26]. As for the B-N system, the existence of subnitride $B_6N$ with $B_6O$-like structure has been suggested based on the experimental data [27]; however, the later critical analysis of that results has shown that the proposed structure (and probably composition) is inconclusive [28]. Our additional *in situ* studies allowed to clarify the situation and will be described below with the example of subnitride $B_{13}N_2$.

Boron also forms metal borides with structures non-related to elemental boron (i.e. without $B_{12}$ icosahedra). Reaction of transition metals with boron (e.g. ambient pressure synthesis of $WB_4$ [29] $ReB_2$ [30]) has been also considered as a potential methodology for obtaining new superhard materials, although generally their hardness do not pass over $H_V$ ~40 GPa, and some high reported values of hardness have been criticized [31]. Reaction of nitrogen with metals is another route (e.g. $Re_2(N_2)(N)_2$ [32] and other compounds [33]), but it is complicated experimentally because of the necessity to make react $N_2$ gas at extreme ($p$, $T$) conditions. Rendering such compounds superhard – for example by nanostructuring – is quite interesting from the fundamental point of view, especially for reaching the ultra-hard limit of $H_V$ ~80 GPa. However the very high density and price remain the main inconveniences of these materials as superabrasives [31]. At the same time, other properties of transition metal borides together with its hardness give a potential for alternative advanced applications. To this end, nanostructured boride precursors can be available by soft chemistry roots [34].

Filled 3-D frameworks like polymerized fullerenes, C- and C-B clathrates are also a challenging group of compounds, produced by hard rigid covalent cages and intercalated metals assuring interesting electronic properties. For example, polymerized and/or partially decomposed fullerenes [35, 36] are superhard, and once intercalated with some metals can show high-temperature superconductivity [37]. Similar polymerization at high pressure has been predicted for carbon nanotubes [38].

Extreme hardness is not only interesting by itself, but also can be intrinsically coupled with other useful properties. For example, clathrate structures are another alternative that can be found in tetrahedral systems, and it is natural to suggest the existence of carbon clathrate frameworks [37, 39]. High hardness comparable to diamond and advanced electronic properties are also expected, including high-temperature superconductivity up to 77 K [40]. In this case high hardness is not only promising for industrial applications, but also from a fundamental point of view. In fact, the light elements like carbon allow creating the rigid covalent bonds in the material, responsible for high-frequency phonon modes (increase of average phonon frequency $\omega_{ln}$) that may significantly increase the critical temperature of superconducting transition $T_c$ in some materials, according to the proportionality $T_c \sim \omega_{ln}$ [40]. Thus, the "carbon framework" compounds are very promising for discovery of superhard superconductors with record $T_c$ and advanced mechanical properties.

Recently the thermodynamic stabilization of $C_3B_3$ cages that can host Sr atoms in the "type-VII" clathrate structure has been predicted [41] with also superconductivity at ~42 K, while in the case of pure carbon frameworks, most of clathrates remain metastable [39]. This suggests the necessity of advanced chemical routes (especially new chemical precursors) to explore such compounds. Here, boron substitution aids in the stabilization of $SrB_3C_3$ clathrate, and offers valuable insights into design guidelines for various carbon-based materials.

### III. The three methodological strategies.

In these most promising chemical systems, three methodological strategies are used to find new bulk super- and ultrahard materials. These are: (1) high pressure allotropy/polymorphism of known elements/compounds, by analogy to the graphite-to-diamond transformation [4]; (2) high pressure polymorphism/synthesis of new dense compounds with unusual compositions; (3) high pressure-induced nanostructuration. In the following, we explain these strategies by illustrating them with results of our recent studies.

### A. High pressure allotropy/polymorphism of known elements/compounds.

This methodology is quite immediate and is the most widely used by research groups as proved by the synthesis of diamond [4, 5] and cubic boron nitride [42, 43]. Actually, the transition from an ambient pressure phase towards a high-pressure (HP) phase, denser, favors the possibility to obtain (if this HP phase is recoverable at ambient conditions) a new superhard material (since the high-pressure phases are often harder than the low-pressure phases of the same compound). This methodology is still used, and even pure elements have recently revealed astonishing surprises: for example, in 2008 the discovery of a new high-pressure phase of boron, orthorhombic $\gamma$-$B_{28}$ (subscript number indicates the number of atoms per unit cell) [23, 44, 45] has completely changed the concept of boron allotropism under pressure [46] and given rise to synthesis of unexpected (and recoverable) boron-rich high-pressure phases [47, 48].

Until 2008, among 16 allotropes of boron mentioned in the literature, only three phases seemed to correspond to the pure element, namely, $\alpha$-B$_{12}$ rhombohedral low-temperature, $\beta$-B$_{106}$ rhombohedral high-temperature low-pressure, and t-B$_{192}$ tetragonal high-temperature high-pressure phases [46]. Discovery of a new high-pressure phase of boron, orthorhombic $\gamma$-B$_{28}$, has been made when studied phase transformations in crystalline $\beta$-B$_{106}$ in a multianvil press at pressures from 12 to 20 GPa and temperatures from 1800 to 2000 K, while its structure has been established by coupling the experimental and theoretical methods, i.e. by powder X-ray diffraction (XRD) measurements and *ab initio* calculations [45]. Later studies on micron-sized single crystals of $\gamma$-B$_{28}$, grown from boron solutions in metal melts at 12 GPa, confirmed the crystal structure of this phase [49, 50].

At ambient conditions $\gamma$-B$_{28}$ has an orthorhombic structure (space group *Pnnm*) with unit cell parameters $a$ = 5.054(2) Å, $b$ = 5.612(3) Å and $c$ = 6.966(5) Å [45]. The structure can be represented as a NaCl-type arrangement of two types of boron clusters, B$_{12}$ icosahedra and B$_2$ pairs (Fig. 2). Although the chemical bonding in $\gamma$-B$_{28}$ is predominantly covalent, there is significantly high charge transfer (of about 0.5 e$^-$) from B$_2$ to B$_{12}$ clusters [45] that is very unusual for elemental crystals. Very recently the nature of chemical bonding in $\gamma$-boron has been explicitly discussed by Macchi [51].

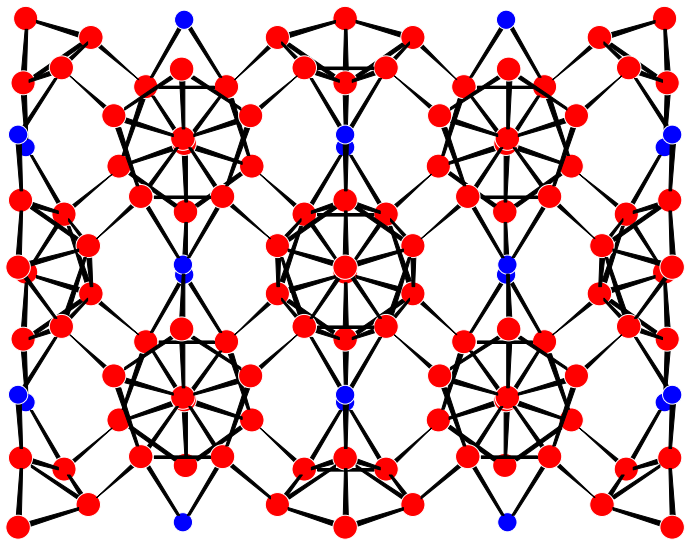

Fig. 2    Crystal structure of $\gamma$-B$_{28}$ [45]. Two oppositely charged sublattices are marked by different colors (cationic, blue; anionic, red).

High hardness often goes together with low compressibility of new phases or, which is the same, by the high bulk modulus $B_0 = V_0(\partial p/\partial V)$ (Fig. 3) [52]. The inverse is not always the case, especially for heavy elements [32], but generally valid for light elements considered in this tutorial. The 300-K equation of state of polycrystalline $\gamma$-B$_{28}$ has been measured in the neon pressure medium up to 65 GPa using a diamond anvil cell (DAC) and synchrotron powder X-ray diffraction [53]. Experimental value of $B_0$ (237 GPa) well agree with theoretical calculations at the DFT-GGA level ($B_0$ = 241 GPa) and single-crystal X-ray diffraction experiments up to 40 GPa ($B_0$ = 237 GPa) [54]. These results allow ascribing $\gamma$-B$_{28}$ to the densest and least compressible form of elemental boron [55].

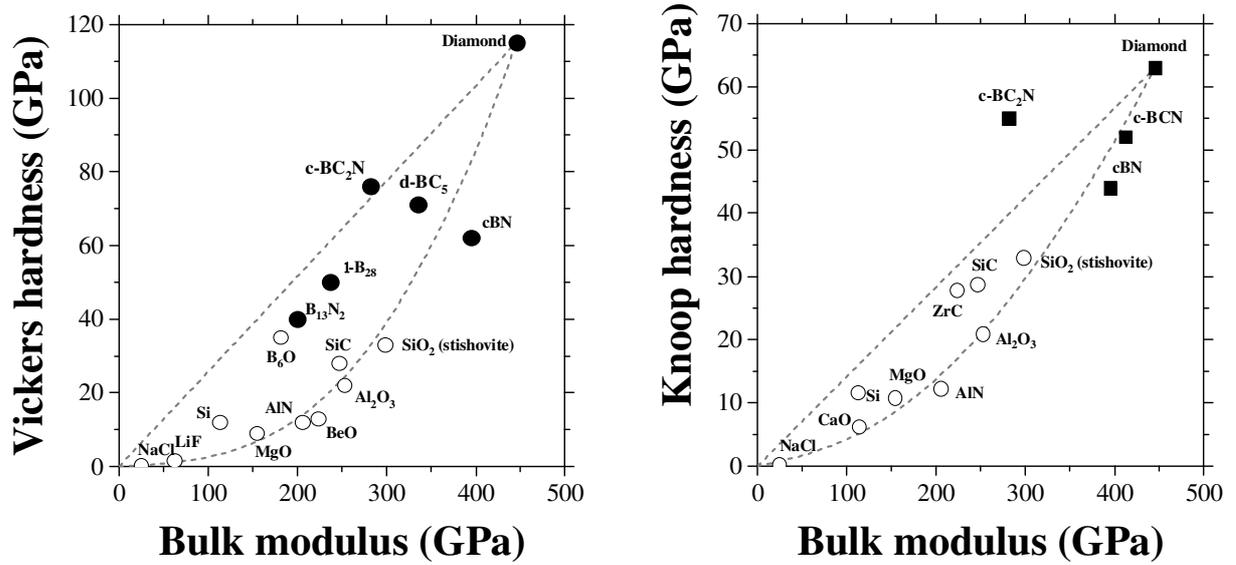

Fig. 3 Vickers (a) and Knoop (b) hardness versus bulk modulus for super- and ultrahard phases of the B–C–N–O quaternary system (black symbols) and some other materials. Grey dashed lines are the guides for eye: linear ($H_V \sim B_0$, $H_K \sim B_0$), cubic ($H_V \sim B_0^3$) and square ($H_K \sim B_0^2$) decencies that delimit most of known materials.

Bulk polycrystalline samples of γ-B$_{28}$ have Vickers hardness of 50 GPa [44] (Fig. 3), which is higher than the hardness of other boron allotropes and agrees well with the 48.8 GPa value calculated in the framework of the thermodynamic model allowing scaling the Vickers hardness values for materials with different bonding types [2].

The discovery of γ-B$_{28}$ provided the missing piece of a puzzle of the phase diagram of boron [45, 56] (Fig. 4). The thermodynamic stability region of this phase is larger than those of all known boron allotropes combined. However, the upper pressure limit of γ-B$_{28}$ stability remains to be studied. Theoretical predictions of an α-Ga-type metallic phase above 74 GPa [57] were confirmed by Oganov et al. [45], except that pressure of this phase transition was shifted to a higher value, 89 GPa, by the presence of γ-B$_{28}$. Recently, this predicted nonicosahedral boron allotrope has be obtained by laser heating of single crystals of β-B to over 2100 K at pressures above 115 GPa [58].

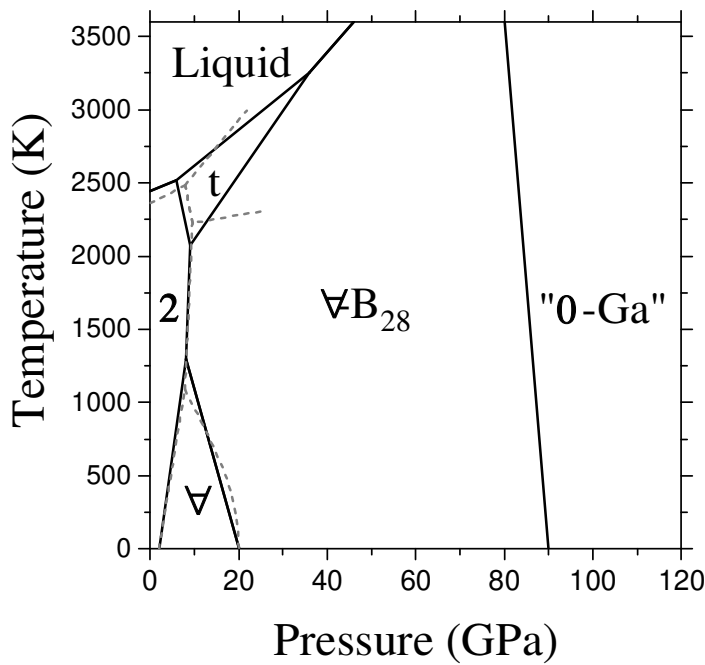

Fig. 4    First tentative *ab initio p-T* phase diagram of boron is represented with solid
lines according to [45], where **α**, **β** and **t** are α-$B_{12}$, β-$B_{106}$ and t-$B_{192}$ boron
allotropes, respectively. Dashed grey lines show refined equilibrium lines
according to experimental observations and thermodynamic analysis [56].

Thus, this methodology allowed the synthesis of a new high-pressure phase of boron, γ-$B_{28}$ which
has the highest hardness [59] and bulk modulus [53] among the known boron crystalline allotropes and
is predicted to be stable in a wide (*p*, *T*) range. The discovery of this boron allotrope intensified the
research in the field and led to a significant number of metastable pure boron forms that could be
synthesized under high pressure and high temperature (HPHT) conditions [47, 48, 60].

## B. High pressure polymorphism/synthesis of new dense compounds with unusual compositions.

The second methodological strategy to synthesize new bulk superhard materials is to explore the
high-pressure polymorphism/synthesis of new dense compounds with unusual compositions. In
practice, this methodology may mean pursuing two distinct paths: (i) to synthesize low-density
precursors of unusual stoichiometry in order to transform them under high pressures to new dense
polymorphs; (ii) to induce chemical reactions under high pressure and high temperature using
conventional precursors to form new dense ultrahard compounds with new compositions. Below we
present these two paths of high-pressure synthesis of new superhard materials via phase transformation of non-standard precursors (example of $d$-BC$_5$) and via a reaction of well-known compounds under extreme conditions (example of B$_{13}$N$_2$). These two pathways are not exclusive, and the last example ($c$-BC$_2$N) illustrates these two possible synthesis paths of the same new compound via the second methodology.

Prior to describing these experimental achievements, we would like to point out one important feature that often raised a controversy in superhard community in the past. Typically, a graphite-to-diamond like phase transformation occurs by combined diffusional and displacive (martensitic) mechanism: high pressure favor displacive, while high temperature - diffusive. Individual mechanisms are possible only in the case of pure C or BN. Such particularity, in the general case of B-C and B-C-N precursors [7, 61] leads unavoidably to nanostructured character of forming phase, with grainsize typically ~5-50 nm. In the terms of a diamond-like structure of B-C-N system, this size range corresponds to 20-200 stacked "buckled" layers along hexagonal c-axis of diamond ([$111$] direction of cubic lattice). To define nanoparticles (individual or aggregated) as unique phase (in thermodynamic sense) is quite challenging problem. Two fundamental notations of phase, i.e. crystal structure and chemical composition, are often questioned, which was the case of ultrahard B-C-N phases that had divided the superhard-material community for decades because of fundamental importance. In fact, the space resolution of typical methods for materials characterization (for example XRD, Raman, high-resolution transmission electron microscopy (TEM) and electron energy loss spectroscopy (EELS)) is often comparable to the grainsize, and systematically only the consistency between different characterization techniques can indicate the chemical homogeneity and individual crystal structure. In this tutorial, we tried to make presentation as smooth and simple as possible, avoiding historical disagreements.

## B.1 Synthesis of diamond-like BC$_5$

Superhard diamond-like BC$_5$ ($d$-BC$_5$ or $c$-BC$_5$) has been synthesized by direct phase transformation of graphite-like B–C solid solutions by analogy of graphite-to-diamond transformation [62, 63]. Experiments at pressures above 25 GPa and temperatures of about 2200 K have been performed in a laser-heated diamond anvil cell. For the comprehensive study of $d$-BC$_5$ by a variety of microscopic characterization techniques, well-sintered ingot of nanocrystalline diamond-like BC$_5$ have been synthesized using multianvil Paris-Edinburgh press [62].

According to XRD powder pattern, the structure of $c$-BC$_5$ is similar to diamond (Fig. 5a). The $d$-BC$_5$ lattice parameter of $a = 3.635(8)$ Å is larger than that of diamond (3.5667 Å), and is close to the 3.646 Å value expected from the ideal mixing (Vegard's law) between diamond and "diamond-like boron" (Fig. 6). The latter hypothetical diamond structure with $a = 4.04$ Å corresponds to the B–B bond length of 1.75 Å [64]. Theoretical predictions also confirm this value [65]. The homogeneity of boron (and carbon) distribution all over the crystal lattice has been confirmed by the EELS elemental mapping (Fig. 5b & c). The valence state mapping also reveals homogeneity of the atomic hybridization (sp$^3$ for both B and C atoms); thus proving the existence of $d$-BC$_5$ as an individual phase [62].

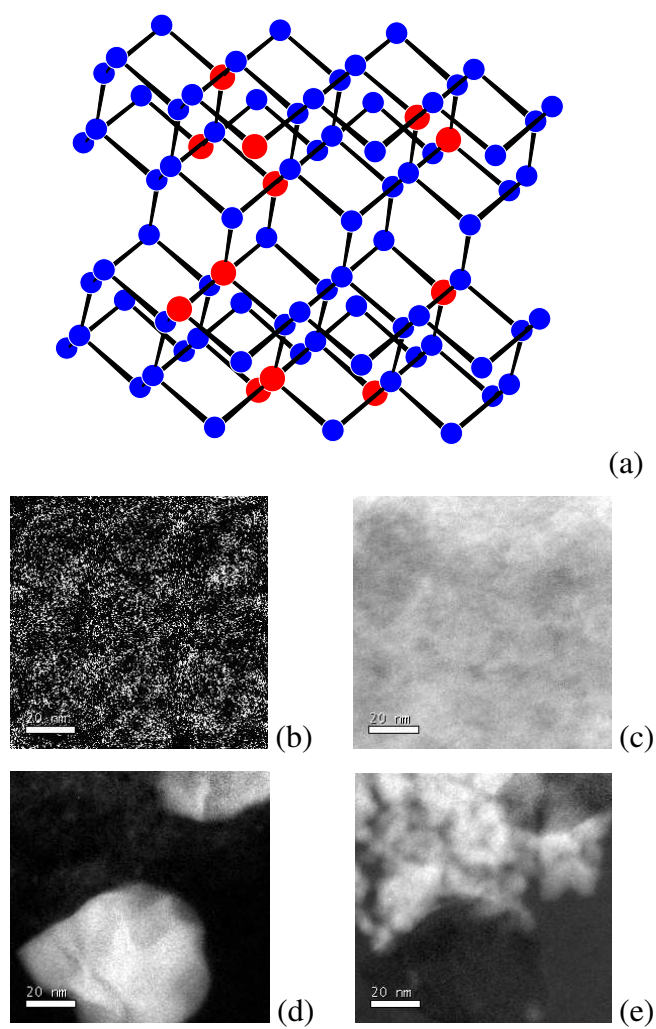

(a)

(b)

(c)

(d)

(e)

Fig. 5   (a) Crystal structure of diamond-like $d$-BC$_5$. The red and blue balls represent the boron and carbon atoms, respectively. The boron atoms are randomly distributed allover the diamond-like lattice. (b-e) EELS element mapping of $d$-BC$_5$ sample before and after thermal decomposition (the scale bar represents 20 nm): B in $d$-BC$_5$ (b); C in $d$-BC$_5$ (c); B in B$_4$C/diamond composite (d); and C in B$_4$C/diamond composite (e).

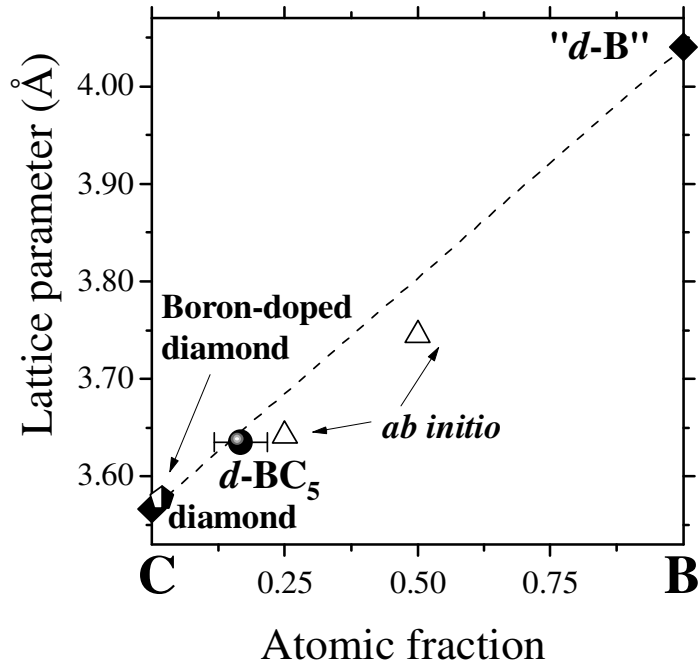

Fig. 6    Lattice parameters of boron substituted diamonds versus boron content. The dashed line represents Vegard's law, while solid circle with error bar shows the value for diamond-like $BC_5$ in comparison with boron-doped diamond with 1.8 at% B (pentagon) [62] and *ab initio* GGA data (triangles) [65].

The *d*-$BC_5$ phase forms in a relatively narrow (~200 K) temperature range and has a metastable character. In fact, it is an ultimate metastable solid solution of boron in diamond, never achieved by the equilibrium dissolving of boron in carbon. Even slight overheating leads to the phase segregation into more thermodynamically stable boron carbide $B_4C$ and boron-doped diamond (see, for example, EELS element mapping at Fig. 5d & e). At the same time, higher thermal stability has been observed at ambient pressure as compared to polycrystalline diamond.

*In situ* XRD measurements of the 300-K EOS up to 40 GPa, allowed obtaining the bulk modulus of $B_0 = 335$ GPa [62]. Such high value is exceeded only by the bulk moduli of diamond (446 GPa [66]), *c*-BN (395 GPa [67]) and some diamond-like $BC_xN$ (see chapter IV), therefore, allow suggesting the very high hardness of diamond-like $BC_5$, as soon as it is produced by light elements and having 3-D diamond structure (possibly, also influenced by nanostructuring).

The obtained ingot material has been tested by various microscopic techniques in order to reveal its technological capacities. The material has similar order of magnitude of electrical conductivity as other doped semiconductors (~0.6 Ω·m at 300 K) and occurs as nanocrystalline aggregates with clearly visible but very small grains with an average size of 10-15 nm. Intrinsic electrical conductivity of superhard material is a highly desired property that makes possible shaping the material by electro erosion, a method of choice that cannot be often applied to superhard materials.

The Vickers hardness was estimated as $H_V = 71$ GPa in the asymptotic-hardness region (recommended for hard and brittle materials [1]) (Fig. 3a). This value is in excellent agreement with the values predicted in the framework of the thermodynamic model of hardness, i.e. 70.6 GPa [52, 68] as well as by microscopic hardness model, i.e. 70 GPa [69]. Under highest applied loads of Vickers indenter (10-N and 20-N), the cracks have been observed and the value of fracture toughness was estimated as $K_{IC} = 9.5$ MPa m$^{1/2}$. Such high value (typically 1 to 5 for various ingot materials produced by polycrystalline diamond and $c$-BN) is indicative of high crack resistance, which can be explained by the blocked crack due to the nanostructuring. Alternative evaluation of hardness by nanoindentation measurements (so-called nanohardness) also have confirmed the superhard character of the phase with corresponding value of 73 GPa [70].

The thermal stability up to 1890 K of bulk diamond-like BC$_5$ ("diamond-to-graphite" structural transition with decomposition) has been measured by *in situ* by XRD in non-oxidizing environment. It is by 500-K higher than thermal stability of nanocrystalline diamond with the same grainsize. This may be attributed to the higher activation barrier for graphitization and/or phase segregation of diamond lattice due to the presence of boron. At higher temperatures, $d$-BC$_5$ decomposes into disordered graphite and amorphous boron and/or boron carbides [62].

Finally, synthesized diamond-like BC$_5$ gives an excellent example of material with extreme hardness combined with unusually high (for superhard materials) fracture toughness, thermal stability and electrical conductivity. This makes it an exceptional superabrasive superior to diamond for some applications, especially those that may require shaping easily achievable by electro-erosion (contrary to binderless diamond and $c$-BN). Electrical conductivity is also characteristic for ceramic composites obtained by decomposition of graphitic BC$_x$ phases at HPHT conditions [71].

## B.2 SYNTHESIS OF BORON SUBNITRIDE B$_{13}$N$_2$

The systematic *in situ* studies of the B–BN system allowed discovery of rhombohedral boron subnitride B$_{13}$N$_2$. Until 2007 it was a missing boron-rich compound of α-boron type of the second period elements [72, 73], and it was also expected to be superhard, similarly to boron carbide B$_4$C [74]. According to the *in situ* synchrotron X-ray diffraction, this phase crystallizes from the B–BN melt at pressures up to 5.3 GPa and temperatures up to 2800 K. It forms brilliant dark red crystals, easily distinguishable in the black boron matrix. The detailed characterization using powder X-ray diffraction, Raman spectroscopy, high-resolution transmission electron microscopy (TEM) and electron energy loss spectroscopy (EELS) showed that the crystal structure of B$_{13}$N$_2$ belongs to the $R$-3$m$ space group and represents a new structural type produced by the distorted B$_{12}$ icosahedra linked together by N–B–N chains and inter-icosahedral B–B bonds [75] (Fig. 7).

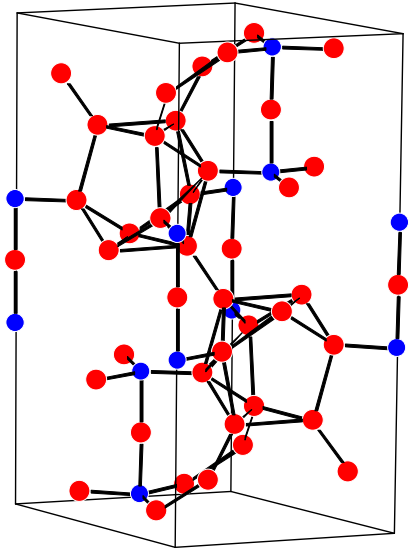

Fig. 7    Crystal structure of rhombohedral $B_{13}N_2$ [75]. The red and blue balls represent the boron and nitrogen atoms, respectively.

In contrast to boron carbide, boron subnitride is a stoichiometric compound and not a solid solution. The Rietveld refinement of XRD powder diffraction data using the $B_4C$-like unit cell as a starting model, gives the composition $B_{13}N_2$ [75]. The site occupancies of atoms of each type are close to unit, so the synthesized $B_{13}N_2$ phase is stoichiometric, in contrast with O-deficiency of $B_6O$ of the same structural type. Besides, the formation of two other boron-rich B–N phases denoted as "$B_6N$" and "$B_{50}N_2$" has been observed. [72, 73] Their structures have not been resolved in original works.

Solozhenko et al. have found that $B_{13}N_2$ is thermodynamically stable boron subnitride, and at 5 GPa it melts incongruently at 2600 K and forms eutectic equilibrium with boron [76]. The equilibrium phase diagram of the B–BN system at 5 GPa (Fig. 8) is characterized by the following nonvariant equilibria:   $L + BN \rightleftarrows B_{13}N_2$ of peritectic type at 2600 K;   $L \rightleftarrows \beta\text{-B} + B_{13}N_2$ of eutectic type at 2300 K;  and $L \rightleftarrows \beta\text{-B} + BN$ metastable eutectic at 2120 K that assures the appearance of the liquid phase, from which $B_{13}N_2$ crystallizes. The evolution of phase diagram of B-BN system up to 24 GPa has been studied in the later work [77]. There are two thermodynamically stable boron subnitrides in the system i.e. rhombohedral $B_{13}N_2$ and tetragonal $B_{50}N_2$. Above 16.5 GPa, the $B_{50}N_2 \rightleftarrows L + B_{13}N_2$ peritectic reaction transforms to the solid-phase reaction of $B_{50}N_2$ decomposition into tetragonal boron ($t'$-$B_{52}$) and $B_{13}N_2$, while the incongruent type of $B_{13}N_2$ melting changes to the congruent type only above 23.5 GPa.

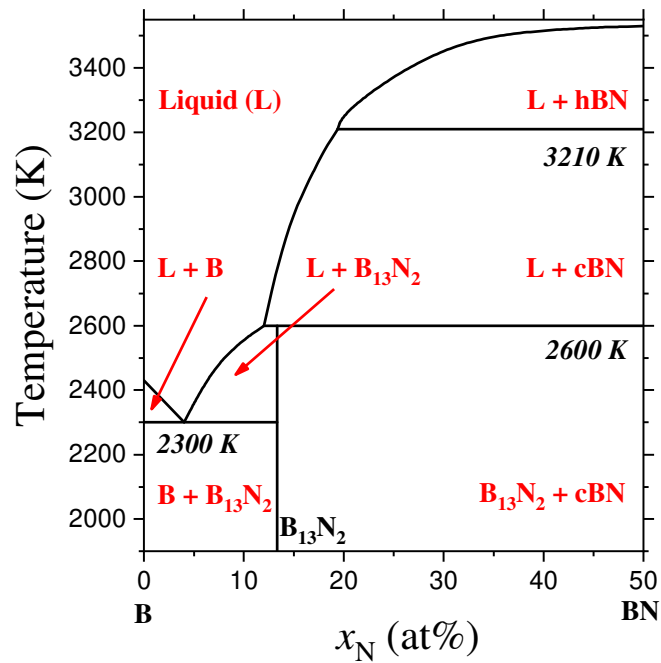

Fig. 8  Phase diagram of the B–BN system at 5 GPa [76] constructed by combination of *in situ* experimental data and thermodynamic calculations.

The 300-K equation of state of $B_{13}N_2$ has been measured up to 30 GPa in a DAC in neon pressure medium using synchrotron powder X-ray diffraction [78]. The value of bulk modulus, $B_0 = 200$ GPa, is close to those of superhard boron compounds such as suboxide [79] and boron carbide [74].

According to the predictions made by Mukhanov et al. [2] in the framework of the thermodynamic model of hardness, the $B_{13}N_2$ subnitride is expected to exhibit microhardness $H_V$ of 40 GPa (Fig. 3) comparable to that of commercial polycrystalline *c*-BN. The experimental measurements of hardness, elastic properties and fracture toughness clearly indicate that $B_{13}N_2$ belongs to a family of superhard phases [80] and can be considered as a promising superabrasive or binder for cubic boron nitride. The experimental value of Vickers hardness $H_V$ of 41(2) GPa [80] is in good agreement with predictions.

### B.3  Synthesis of superhard cubic $BC_2N$

Superhard cubic $BC_2N$ (*c*-$BC_2N$) has been first synthesized in 2001 and became second to diamond superhard material filling the hardness gap between cBN and diamond. Its discovery gave a renaissance to modern intense research of superhard phases and reinforced the idea of graphite-to-diamond-like structural transformations as a route to extreme hardness. By direct solid-state phase transition of graphite-like $(BN)_{0.48}C_{0.52}$ [81], *c*-$BC_2N$ has been synthesized at pressures above 18 GPa and temperatures higher than 2200 K using laser-heated DAC [82]. Segregation of graphitic B-C-N compound into cBN and diamond (or disordered graphite) has been observed at lower pressures.

The macroscopic samples (up to 2 mm$^3$) have been produced using a large-volume multianvil apparatus with a 5000-ton press at Bayerisches Geoinstitut.

$c$-BC$_2$N has cubic diamond structure (and can be named $d$-BC$_2$N using the notations of this tutorial) that corresponds to the Fd-3m space group. Powder diffraction pattern shows only *111*, *220*, and *311* Bragg reflections, while the absence of the *200* line (typical for $c$-BN) indicates that B, C and N atoms are uniformly distributed over both *fcc* sublattices of the zinc-blende structure. The lattice parameter of $a = 3.642(2)$ Å [82, 83] is noticeably larger than those of both diamond (3.5667 Å) and cubic boron nitride (3.6158 Å) (Fig. 9). Thus, the synthesized phase cannot be considered as "diamond–$c$-BN alloys" with lattice parameters that follow ideal mixing law [84].

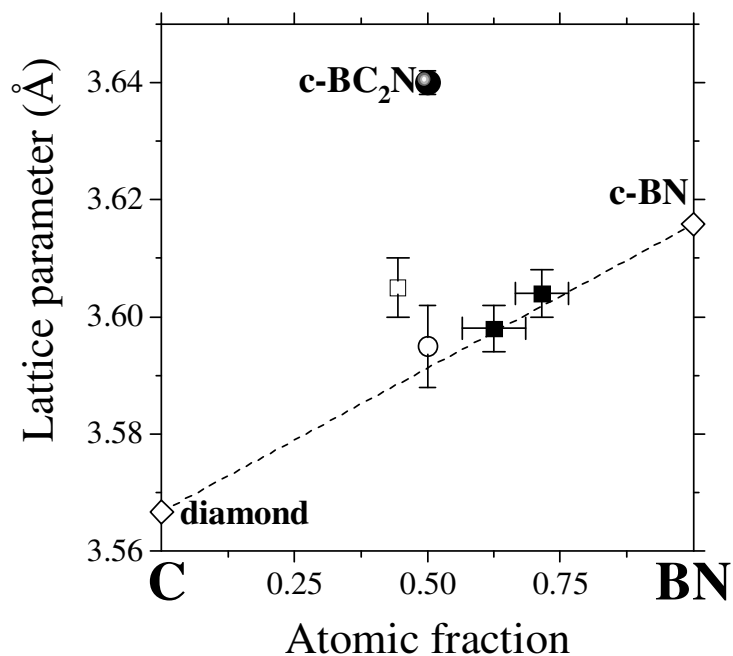

Fig. 9    Lattice parameters of cubic BC$_2$N [82] (solid circle), zinc-blende BC$_2$N [85] (open circle), BC$_{2.5}$N heterodiamond [86] (open square) and diamond-like BN-C solid solutions [84] (solid squares). The dashed line represents ideal mixing between diamond and cubic boron nitride.

Analytical TEM showed the grain size of $c$-BC$_2$N ranges from 10 to 30 nm [87], which is typical for metastable diamond-like phases produced by "graphite-to-diamond-like" transformation (see also $d$-BC$_5$ example above). Selected area electron diffraction patterns confirm the absence of superstructure and uniform distribution of B, C, and N atoms in the diamond lattice. The $K$-edge EELS spectra of B, C and N atoms, which are sensitive to orbital hybridization, revealed an $sp^3$ type atomic bonding and additionally confirmed the formation of diamond-like B–C–N phase [87]. The atomic force microscopy studies of granular structure of bulk $c$-BC$_2$N indicated the average size ~200 nm [88]. This large grains are aggregates of small crystallites of 20-30 nm.

Raman study using both UV and visible excitation lights [89] indicated that the position of the unique Raman band (at 1325.7(16) cm$^{-1}$) is located between the Raman peaks of diamond and $c$-BN, and corresponds to LO mode of $c$-BC$_2$N phase. The large full width at half maximum (FWHM) of this band (~25 cm$^{-1}$) most likely reflects random atomic distribution all over the diamond lattice.

While the lattice parameter of $c$-BC$_2$N is larger than that expected from Vegard's law applied to cBN and diamond, the bulk modulus of $B_0 = 282$ GPa [82] is ~30% smaller than the 420 GPa value expected from ideal mixing. The *in situ* XRD measurement in DAC to 30 GPa [82] has been later confirmed by the value obtained by Brillouin scattering [90]. The bulk and shear moduli of $c$-BC$_2$N have been estimated at 259 GPa and 238 GPa, respectively [90]. The shear modulus of 447 GPa evaluated earlier [82] from the load-displacement curves is most likely an overestimate due to distinct deformation of the diamond indenter during nanohardness measurements.

Micro- and nanoindentation techniques allowed characterization of mechanical properties of $c$-BC$_2$N [82]. The values of nanohardness ($H_N = 75$ GPa), as well as Vickers ($H_V = 76$ GPa) and Knoop ($H_K = 55$ GPa) microhardness are always intermediate between those of diamond and $c$-BN. that makes cubic BC$_2$N the second hardest known material (Fig. 3). The elastic recovery of $c$-BC$_2$N has been found to be 68% which is higher than the corresponding value for $c$-BN (60%), and is approaching that of diamond. In fact, $c$-BC$_2$N phase has an unusual combination of mechanical properties: its elastic moduli measured by Brillouin scattering and X-ray diffraction are lower than those of cubic boron nitride, whereas its hardness measured independently by micro- and nanoindentation techniques is higher than that of single-crystal $c$-BN and is only slightly lower than that of diamond.

The *in situ* study of thermal stability have shown that $c$-BC$_2$N remains stable up to 1800 K, which makes it more stable than polycrystalline diamond with the same grain size. At pressures of 25-32 GPa, the decomposition into $c$-BN and diamond has been observed at temperatures above 2900 K [83].

Distinctive dense polymorph of BC$_2$N, nanostructured bulk zinc-blende BC$_2$N ($zb$-BC$_2$N) material, has been obtained from ball-milled mixture of graphite and hexagonal boron nitride at 20 GPa and 2200-2250 K [85]. The well-sintered translucent chunks had the Vickers hardness of 62 GPa. According to the high resolution TEM, the obtained $zb$-BC$_2$N is nanocrystalline with a grain size of ~5 nm with amorphous grain boundaries. Contrary to diamond-structure $c$-BC$_2$N described above, crystalline $zb$-BC$_2$N is consistent with a face-centered cubic zinc-blende structure. The lattice parameter of $a = 3.595$ Å well agrees with ideal mixing law, which renders $zb$-BC$_2$N significantly different from $c$-BC$_2$N synthesized in [82] (Fig. 9).

To shed some light on distinctive forms of BC$_2$N, their crystal structure features and to understand quite significant spread in properties, first-principles calculations have been performed [91]. The proper choice of the supercell, cutoff energy and sampling $k$ points allowed to show the stability of cubic phases. The bulk moduli for the phases with different atomic distributions all over the supercell (e.g. high-density phase with C-B-N layered superstructure, high-density phase without any C-B-N layers, and low-density phase; all the phases are defect-free and do not possess any B–B or N–N bond) were in excellent agreement with experimentally established values. The computational results conclude that the low-density phase, $c$-BC$_2$N synthesized by Solozhenko et al. [82], has no C–C bonds, whereas the high-density phase, $zb$-BC$_2$N synthesized by Zhao et al. [85] has C–C bonds. The unique feature of each of the cubic BC$_2$N phases is, therefore, a result of the local electronic structure and chemical bonding in the crystal structure.

Later studies on elastic moduli and strength of nanocrystalline $zb$-BC$_2$N under nonhydrostatic conditions up to 100 GPa [92] have shown that the compound could support a maximum differential stress of 38 GPa when it started to yield at about 66 GPa under uniaxial compression.

## C. High pressure-induced nanostructuration

Finally, the last methodological strategy to synthesize new "bulk" super- and ultrahard materials relies on the high pressure-induced nanostructuration. Actually, according to experimental observations and theoretical simulations, for the majority of polycrystalline materials a decrease in grain size down to dozens of nanometers results in significant increase in hardness compared to single crystals and polycrystalline (microstructured) materials, up to 70% in some cases [9, 42, 93-96]. This phenomenon is known for a long time and is called the Hall-Petch effect (HPE). Here, for the clarity, we will present only simplified model of this phenomenon, while more mechanisms (like twinning) have been previously discussed in other review.[97] Anyway, the role of high-pressure synthesis is just the same as for new composition and crystal structure stability: tunable nanostructuration by playing p-T-time conditions of synthesis.

The HPE can be explained schematically using the concept of dislocation pile-up: in polycrystalline materials, grain boundaries are barriers to dislocation motion, notably because of the crystallographic mismatch between adjacent grains that requires more energy for a dislocation to change directions and move into the adjacent grain. Hence, under an applied stress, dislocations concentrate in a grain until dislocation sources are activated in the neighboring grain, allowing further deformation in the material. Hence, reducing the grain size means that the number of grain boundaries per volume unit increases and that, for each grain, the number of dislocations that can pile-up is reduced. Therefore, the amount of applied stress needed to move dislocations across the grain boundaries is increased compared to coarse-grained materials: it means that the hardness is increased.

Thus, there is an inverse relationship between grain size $d$ and yield strength $\sigma$ that is described by the Hall-Petch equation (Eq. 1). Both $k$ and $\sigma_0$ are material-dependent parameters, namely, the strengthening coefficient and the friction stress in the absence of grain boundaries, respectively. They characterize the resistance of the lattice to dislocation motion.

$$\sigma = \sigma_0 + \frac{k}{d^2} \qquad (1)$$

However, this equation is not valid for all grain sizes [98]. As plotted in Fig. 10, when grains are smaller than the critical value $d*$ (typically around 10 nm), the hardness reduces, causing the "inverse Hall-Petch effect". To resume: above $d*$, as the grain size is reduced, less and less dislocations can pile-up in a grain, increasing the overall hardness. When the grain size is $d*$, grains can accommodate only one dislocation, meaning the maximum of strengthening is reached. Hence, if the grain size is further decreased, below $d*$, other yielding mechanisms may come in play. If they are not entirely understood yet, recent studies suggest that Coble creep and other diffusional phenomena are to be taken into account [98].

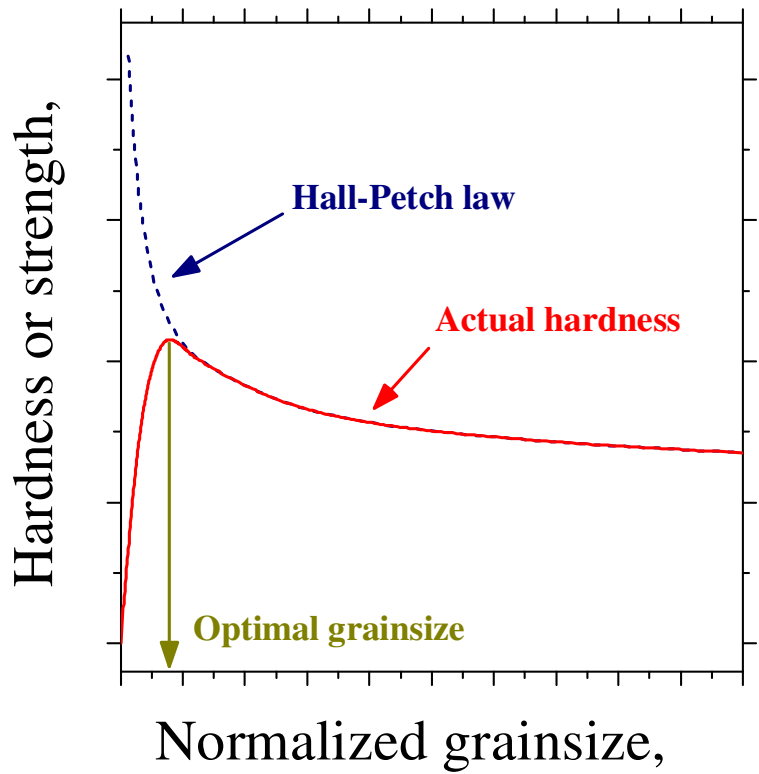

Fig. 10  Material's hardness as a function of grain size.

The critical grain size $d*$ would then correspond to a shift in the dominated plastic behavior from dislocation-mediated plasticity to grain-boundary-associated plasticity (sliding, grain boundary diffusion, etc.) as dominant fracture propagation mechanisms. Pande and Cooper [98] proposed then to generalize the yield stress applied to a polycrystalline material according to Eq. 2 :

$$\sigma = \sigma_0 + \frac{k}{d^2} + k_1 + \frac{B_0}{d} + B d^3 \tag{2}$$

The last three terms describe the diffusion-based dislocation transport. $B_0$ and $k_1$ are constants and $B$ is a strain and temperature dependent parameter. For large grain size, the first two terms are dominant, according to the Hall-Petch effect. For small grain sizes, the last three terms become dominant, leading to the inverse Hall-Petch effect.

To resume, the strength/hardness of a polycrystalline material has been found to increase with decreasing grain size down to a critical nanoscale value (usually about 10–20 nm). This is particularly interesting in the case of bulk hard or superhard polycrystalline materials as it can yield improved mechanical properties and performances. That's why the synthesis of bulk nanostructured ultrahard materials remains a very challenging domain. The common methods of soft chemistry allow obtaining nanoparticles, for example nanodiamond [99-106], whose direct sintering, at high temperature, usually leads to the grain growth and loss of nanostructure. To solve this problem, one idea is to reduce drastically the temperature and the time of sintering, but in the literature, many works have been reported showing that this idea to sinter nanodiamond powders to produce a bulk polycrystalline diamond fails. The authors always observed unavoidable graphitization, high porosity, sometimes impurities, heterogeneous stress distribution, and also poor intergrain adhesion [107]. And, of course, in polycrystals, grain boundary cohesion is a crucial factor influencing hardness.

In fact, obtaining bulk nanopolycrystalline ultrahard materials can be achieved by fortunate combination of, at least, two main parameters: a good choice of starting material (purity, microstructure, etc.) and the synthesis conditions, which means pressure, temperature and duration of synthesis. These latter parameters are crucial because combining extreme ($p$, $T$) conditions are powerful and promising tools for grain-size control during direct solid-state phase transformations. Actually, the simultaneous increase of pressure and temperature make possible to combine different nucleation, growth and aggregation regimes (also very important because responsible of intergranular bonding strength) with high flexibility, and, therefore, to go deep into nanoscale engineering. Schematically, high pressure and moderate temperature is the best combination for pressure-induced nanostructuration in solid-solid transformation as high pressure increases the driving force of transformation that favors nucleation, and moderate temperature suppresses grain growth by reducing the atomic diffusion responsible for the diffusion growth of grains (Cf. Fig. 11).

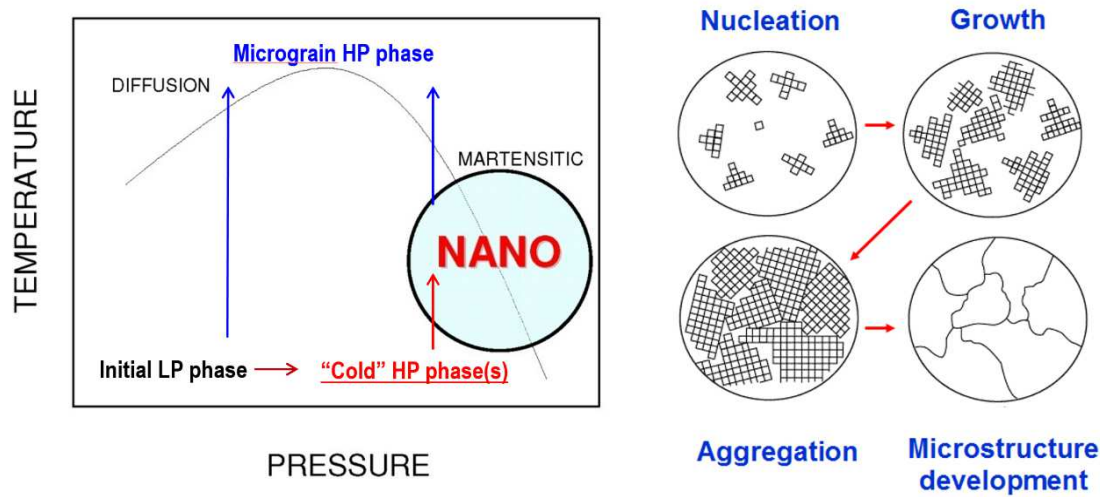

Fig. 11   Schematic diagram showing the ($p$, $T$) domain where different nucleation, growth and aggregation regimes of solid-solid phase transformation can allow nanostructured high pressure phase synthesis. High pressure and moderate temperature increases the driving force of transformation that favors nucleation, and suppresses grain growth by reducing the atomic diffusion responsible for the diffusion growth of grains.

The exploitation of this methodology has been successfully applied to ultrahard materials, first of which diamond, reaching binderless nanopolycrystalline diamond (NPD). Actually, in 2003, Irifune et al. prepared binderless NPD by direct conversion from graphite at 12-25 GPa in the 2300-2500°C temperature range [93]. Very high hardness values and high thermal stability were claimed and more synthetic details were given in a further study [108]. Materials were prepared in a multianvil apparatus, allowing large samples to be recovered and characterized. Knoop hardness measurements were performed on the synthesized samples. At synthesis conditions of 18 GPa, 2500°C for 1 min, transparent pure cubic diamond is obtained with $H_K$ = 100-110 GPa. At synthesis conditions of 15 GPa, 2400°C for 1 min, transparent pure cubic diamond is also recovered with $H_K$ = 130-140 GPa. TEM observations show that the cubic diamond samples consist of 10-20 nm crystals, randomly oriented, as evidenced by electron diffraction [93]. Consequently, it is possible to prepare NPD that possesses hardness at least equal to that of single crystal diamond ($H_K$ = 120 GPa maximum, in the appropriate crystallographic direction), and even higher.

Following the same methodological strategy, similarly to diamond, single-phase (binderless) nano-polycrystalline $c$-BN has been synthesized [109] by optimizing and controlling the starting material and synthesis conditions:

1/ According to the data of mass-spectrometry, electron probe microanalysis and electron energy loss spectroscopy, the starting pyrolytic boron nitride $p$-BN samples contain only boron and nitrogen (in fact, "pyrolytic" is referred to the method of synthesis, i.e. pyrolysis, and can have various layered graphite-like structures, either ordered or disordered along hexagonal $c$-axis [7]). No detectable amounts of chlorine, hydrogen and oxygen have been observed; this high purity and zero porosity of initial bulk pyrolytic boron nitride assured the high purity of resulting nanostructured $c$-BN phase. Perfectly turbostratic structure (zero degree of ordering along c-axis, or random layer lattice [110]) allowed achieving quasi perfect $c$-BN without stacking faults; while in the case of commercial turbostratic BN powders lead to nanocomposites with $w$-BN domains [111].

2/ At 20 GPa, the direct conversion $p$-BN to $c$-BN occurs, and no secondary phases are observed. Also, when the temperature of synthesis decreases, the decrease of the grain size of polycrystalline cBN from 300 nanometers down to 20 nanometers is observed [109]. We consider as nanostructured cubic boron nitride (nano-$c$-BN) a polycristal of cubic boron nitride with grain size at about 20 nm. At 20 GPa, nano-$c$-BN phase can be synthesized only in a very narrow temperature range, between 1770 et 1870 K. The evolution of hardness of as-synthesized nano-$c$-BN samples with their grain size follows the well-known Hall-Petch effect (Cf. Fig. 12). With the grain size decrease, the hardness increases from $H_V$ ~40 GPa for microcrystalline $c$-BN up to 85 GPa for nano-$c$-BN, approaching that of diamond. The combination of nanosize and high purity also led to very high fracture toughness and wear resistance [109, 112].

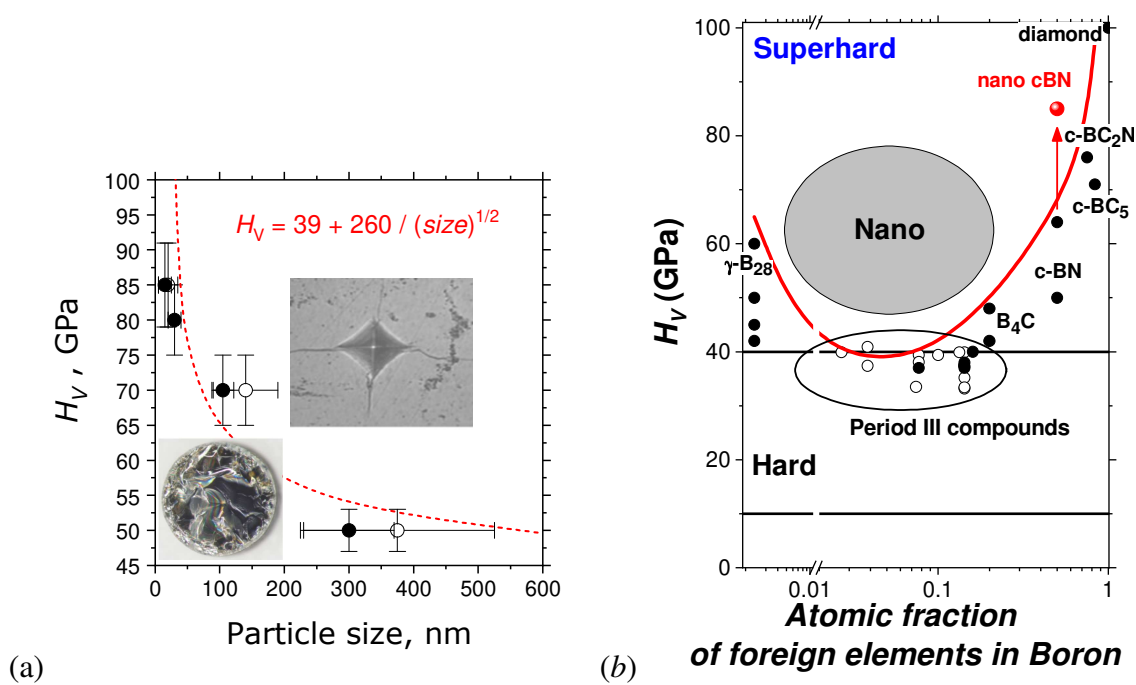

(a)                                                        (b)

Fig. 12   (*a*) Hardness of nano-*c*-BN fitted to the Hall-Petch equation. Inserts show the sample picture and the indentation. (*b*) Hardness of boron compounds as a function of boron content (solid circles show known phases, while open circles - predictions). Red line shows the higher limit of bulk hardness. Nano-structured materials, such as nano-*c*-BN (red circle), allow overpassing this boundary.

These recent examples of high-pressure synthesis of extremely hard bulk binderless nanomaterials based on diamond [93] and cubic boron nitride [109] have also inspired the search for new bulk nanocrystalline materials based on compounds of light elements that may enhance useful properties of their microcrystalline counterparts. This field is still at its initial stage, and a large family of new superhard materials synthesized at high pressure are yet to be discovered with this methodological strategy.

## IV. Experimental and theoretical tools in this field

This section presents the various experimental and theoretical tools in high-pressure synthesis of super- and ultrahard materials so that the non-specialist readers can get a general idea of the studies in this field.

## A. High-pressure tools

To synthesize new bulk super- and ultrahard materials under high pressure, two types of apparatus are mainly used: the diamond anvil cell and the large volume presses.

The invention of the diamond anvil cell (DAC) at the beginning of the 50's [113] was a revolution for the field of static high-pressure science. Diamond is the hardest material and best thermal conductor and therefore is an excellent material for high pressure – high temperature experiments. In addition, diamond is transparent in a very wide spectral range allowing *in situ* spectroscopy and X-ray diffraction of compressed materials giving access to microscopic vibrational and structural information (Cf. below). A DAC is a compact and light (~200 g) piston-cylinder device which can generate pressures of the order of 400 GPa. The pressure is classically measured *in situ* by visible fluorescence or X-ray diffraction of a known standard. Laser heating is commonly used for reaching very high temperature (up to 4000 K in routine). The main disadvantage of the diamond anvil cell is the small sample size (few hundreds of microns at the maximum), therefore it is difficult to recover the sample and to characterize it. The diamond anvil cell is mainly used in this field as a simple tool for *in situ* exploration (Raman or X-ray diffraction) to target areas ($p$, $T$) that we will then reproduce in a large volume cell for complete characterizations. A striking example of this is the $d$-BC$_5$ which was first synthesized in a DAC to precisely determine the synthesis conditions ($p$, $T$) before reproducing this synthesis in a large-volume cell for complete characterization.

Large-volume cells (the history, designs and calibrations of different types of large-volume presses have been discussed in the reviews [114-116]) can be used to compress samples (greater than 1 mm$^3$ in size) to pressures above 1 GPa. In order to reach high temperatures, resistive heating is used through an internal heater such as graphite, rhenium, etc. Thermocouples are also used in tandem to measure the value of the said high temperature. Therefore, it is possible to simultaneously attain high pressure and high temperature in these cells. Multi-anvil systems have been instrumental in devising precise phase diagrams of materials under conditions that exists 800 km below the Earth's surface. They have also proved invaluable for synthesis of many novel and metastable solids. Multi-anvil cells can produce thermodynamic equilibrium states of materials under controlled, known and independently determined pressure and temperature parameters. For a number of years, large-volume high-pressure devices could only be used in the 1-10 GPa range, thus inhibiting the exploration of new materials and phases beyond this limited pressure range. However, technological advancements in the recent past has allowed us to extend the range to 20 GPa and sometimes, even more. Temperatures up to 3000 K can now be attained for the formation of refractory compounds. These new frontiers allow one to synthesize bulk compounds that are impossible to produce by other methods. The new phases such as $d$-BC$_5$ (section III.B), nano-$c$-BN (section III.C) and gamma-boron (section III.A) are striking examples since they all are synthesized at pressures between 10 and 20 GPa.

In addition to the mentioned above high-pressure tools, it is also possible to use dynamic pressures that can be complementary to the static pressures, allowing even more extreme ($p$, $T$) conditions on large recovered volumes. As an example, in the B-C-N system several interesting studies can be mentioned which illustrate the interest of this still ill studied domain.

First shock-compression synthesis of the heterodiamond with $BC_{2.5}N$ stoichiometry has been reported by Komatsu et al. [86] as a product of transformation of graphitic B-C-N precursors in a copper matrix at 50 GPa and 3500 K. For the synthesis, the cylindrical apparatus and AN-TNT explosives were used [117]. The X-ray diffraction studies have shown the lattice parameter of $BC_{2.5}N$ of cubic lattice $a = 3.605$ Å, close to that of cubic boron nitride (3.6158 Å) (Fig. 9). The recovered nanocrystalline powders have grain size of 5-20 nm and oxidation resistance comparable to that of diamond. Hardness measurements were not performed, while the estimated bulk modulus (401 GPa) allowed the authors to suggest that their B-C-N heterodiamond is the hardest material next to diamond.

The shock-compression synthesis of diamond-like B–C–N phases has been also studied by Solozhenko et al. [83, 118]. The incident shock pressures on the samples have been controlled by choosing an explosive composition, while the use of the special additive allowed heating up to 3500 K, and abrupt (~$10^8$ K/s) cooling at decompression. Under such conditions, graphite-like $(BN)_xC_{1-x}$ ($0.48 \leq x \leq 0.61$) solid solutions convert into diamond-like phases, with yield up to 80 wt% at 30 GPa [83]. The stoichiometry of both phases can be assumed to be BCN. Diamond-like BCN has the B, C and N atoms that are statistically uniformly distributed over diamond crystal lattice. Profile analysis of observed diffraction patterns confirm the model of the diamond-like BN–C uniform solid solutions, and cannot be a mechanical mixture of diamond and $c$-BN [118]. Lattice parameters of $c$-$BC_{1.2\pm0.2}N$ and $c$-$BC_{0.9\pm0.2}N$ are 3.598 Å and 3.604 Å, respectively [83]. These are expected values according to an ideal mixing between diamond and $c$-BN (Fig. 9).

A mean grain size of both cubic BCN phases is about 5 nm [83, 87]. TEM also showed that the coarsest grains are of a tetrahedron habit, while fine grains have mainly a round shape. Selected area electron diffraction patterns are fully consistent with diamond-like BN-C solid solutions, without any superstructure reflections, which additionally confirmed random distribution of B, C and N in the diamond crystal lattice. The stoichiometry of $c$-BCN phases was confirmed from the B$K$, C$K$, and N$K$ EELS spectra, and found to be BCN in both cases.

The bulk modulus of $c$-BCN, $B_0 = 412$ GPa [83], is higher than that of $c$-BN (395 GPa [67]), and is close to the 420 GPa value that is expected from ideal mixing between diamond and $c$-BN. This make the diamond-like $c$-BCN solid solution one of the least compressible superhard phases, being second only to diamond (446 GPa [66]).

The ingot of shock-synthesized powder of $c$-$BC_{0.9\pm0.2}N$, sintered at high pressures and temperatures, has Knoop hardness as high as 52 GPa [83], which is only slightly lower than that of cubic $BC_2N$ [82] (Fig. 10).

B. *In situ* studies

Research over the last ten years has seen intensive use of *in situ* synchrotron radiation for direct observation of both stable and metastable super- and ultrahard materials synthesis pathways under extreme conditions. This strategy removes the limitations of the old *ex situ* 'cook and look' procedure, which was rather time-consuming research method. The possibility of observing synthesis *in situ* permits much greater precision in establishing the thermodynamic conditions needed for accessing metastable states. Indeed, only *in situ* studies can address the issues in all their complexity and are crucial to understand the mechanisms and kinetics of high-pressure phase transformation. *In situ* X-ray diffraction under high pressure and high temperature (HP-HT) conditions is useful in a number of ways. It permits us to: (i) see in real time the evolution of the structures of the precursors, (ii) find out the eventual intermediate phases that are produced during the reaction, (iii) get a better understanding of the reaction mechanisms at play, (iv) understand (with further analysis) binary phase diagrams at different pressures, (v) get insights into the local order of solutions under extreme conditions of temperatures and pressures (in particular cases where the reaction occurs in solution), (vi) determine the kinetic parameters and thereby recognize the limiting factors which affect the synthesis, (vii) optimize the thermodynamic pathways leading to synthesis (i.e. minimizing the required ($p$, $T$)parameters), (viii) find out the optimal ($p$, $T$) paths for recovering the phases formed at high pressure and high temperature (HPHT) conditions at ambient conditions.

Several examples showing the importance of *in situ* studies in this area of research have been already given in this tutorial paper, especially for *d*-BC$_5$ or *c*-BC$_2$N (section III).

## C. Theoretical calculations

The *ab initio* calculations are an important complement to conventional high-pressure synthesis techniques. Most traditional *ab initio* methods refer to direct calculation of properties for materials with known crystal structures and known or hypothetical compositions. They are mainly limited atomically ordered systems (one atom per crystal site) and can be efficiently used at 2 levels. On the one hand, they are useful to guide the choice of the most promising systems and ($p$, $T$) conditions for attempts of synthesis by calculations for a set of stable or metastable structures or compositions [24, 26]. On the other hand, they are crucial to assist in the interpretation of experimental data concerning the physico-chemical or structural properties of the new materials synthesized. Typically, it is important to establish the thermodynamic stability of a known HP-HT phase [119]. Another important example is a crystal structure solution assisted by *ab initio* calculations (Cf. section III of this paper and refs. [45, 120]).

In fact, density-functional theory (DFT) calculations have been strongly developed in recent years. Their reliability is well established to study such exotic properties as high-$T_c$ superconductivity in complex oxides and manganates, transport properties in transition metal sulfides, as well as mechanical properties, from well-defined elastic moduli that can be easily derived from the total energy [121] to complex material characteristics such as hardness that may be scaled to a 10% accuracy by semiempirical *ab initio* models considering bond energetics and electron density distribution [122, 123], or by direct stress-strain curve simulations for given *hkl* crystal plane [124-126]. Thus, the desired properties can be also considered like a selection criterion for a structure determination during optimal structure research, not only thermodynamic (absolute minimum of energy) or dynamic (relative minimum of energy or absence of imaginary phonons) stability, corresponding to stable or metastable states.

Numerous structural prediction algorithms based on *ab initio* calculations have been proposed to date. In particular, the Oganov group developed an approach based on evolutionary algorithm USPEX (Universal Structure Predictor: Evolutionary Xtallography) [21], which – outside high predictive power – was already particularly fruitful for discovery and characterization of carbon- and boron-based materials [45, 120]. This method requires no experimental data (with the exception of the chemical formula) and is remarkably efficient and reliable, with reasonable calculation times and practically perfect scalability. USPEX is based on the structure prediction evolutionary algorithm that searches for the structure corresponding to the global minimum of total energy. The quality of trial structures in the terms of free energy is calculated by an external *ab initio* code [21]. Systematic search for novel hard materials, using global optimization algorithms and hardness like optimization criterion is now possible [127]. Without any doubt, in the near future this type of algorithm will be the basis of the design of new advanced materials combining high hardness with other useful properties.

## D. Importance of precursors and better control of experimental conditions

The numerous efforts undertaken in these two fields (quality of the precursors and perfect control of the experimental conditions) explain the quality and sometimes the novelty of the ultrahard materials synthesized under high pressure and high temperature. Mentioned above pyrolytic BN and turbostratic graphite-like *t*-BC$_5$ are good examples, how the design of new precursors can impact the properties of resulting high-pressure phases. Moreover, their formation occurs under specific pressures and in quite narrow temperature range that require precise control of experimental conditions. For example, shock compression of *t*-BC$_5$ with nonflexible pressure-temperature control does not allow obtaining *d*-BC$_x$ similarly to *d*-BC$_5$ [128, 129].

## V. Perspectives

Modern high-pressure synthesis of superhard and, especially, ultrahard materials represents a vast exciting area of research which can lead to new industrially important materials. This field is still at an initial stage, and a large family of new super- and ultrahard materials synthesized at high pressure are yet to be discovered. At this end, if the methodologies remain similar to those mentioned in section III, several improvements of the experimental and theoretical tools (see section IV) will open new horizons.

*Higher pressure*

In the coming years, the pressures achievable in large-volume presses will be greatly extended by using stronger anvil material. From today, by replacing the tungsten carbide with nanopolycrystalline binderless diamond for the cubes of the multianvil devices, pressures of 100 GPa could be reached while preserving appreciable sample volumes. Today's records will become more generalized, and more and more laboratories will be able to access extreme pressures that have been poorly explored until now. It should also be noted that the goal of maximizing the sample volume in the "usual" pressure range is of equal importance since this makes possible to study more complicated systems, thus, increasing the synthesis opportunities.

*In situ studies*

The improvement of X-rays detectors, optics and different upgrades that the 3rd generation synchrotron radiation sources should make it possible to carry out *in situ* experiments that will be more precise and fast. New possibilities opened up by imaging and tomography under extreme conditions [130] will allow better understanding of the processes involved in the synthesis of new super- and ultrahard materials.

*Theory*

Theoretical predictions take an important place in the modern design of advanced materials. Beside described above structural evolutionary algorithm USPEX [21], a number of alternative algorithms will be used to generate crystal structures and help to select the most appropriate $(p, T)$ conditions for the materials synthesis [131].

Artificial intelligence (AI) allows developing a set of models representing the problem [132], e.g. materials' set with a given property (hardness in our case). The initial data set is constructed by the system (or, alternatively, it may be combined with some experimental data), and after the tentative models are produced. The choice of the best between them at each iteration step is made using AI framework. The algorithm may start with as few as two experiments as initial data points, and can optimize the experiments discerning the best path [132]. In fact, the human accumulation of experience is replaced by machine learning.

Finally, simulation has greatly impacted the way to perform research, allowing for virtual experiments at greatly reduced cost and with an unprecedented level of control. Also, in addition of these simulations, data science promises to greatly accelerate the process of scientific discovery. Several materials database initiatives have been pursued in the last years. They combine ab-initio calculations with high-throughput methods in order to simulate a very large number of unique systems and compute several distinct properties. These initiatives have already produced detailed data on millions of various structures, all of which is freely available to the scientific community through open databases. In the future, these databases will be very useful in the search and optimization of super- and ultrahard materials for very specific purposes, radically transforming the way in which materials science progresses.

*Improved precursors for more complex materials*

In the coming years, high-pressure synthesis will benefit from the increasingly richness of the precursors chemistry. New complex materials may be considered thanks to this progress. It will be the case for nanocomposites whose structural organization of the precursor involves that of the final material, synthesized under extreme conditions. For example, recently, the original combination of chemically-derived metal boride nanocrystals [34] with HP-HT treatments was used to prepare innovative nanocomposites by high-pressure crystallization of ternary borates [133]. Nanostructuration was preserved upon crystallization of the matrix at high pressure and high temperature, with the size of the nano-inclusions not exceeding 30 nm at pressures as high as 5 GPa. These results validate this new approach combining solution-phase synthesis of inorganic nanomaterials and high pressures to yield nanocomposites made of phases not reachable at ambient conditions. This approach paves the way to advanced ultrahard materials which could combine the best functional properties, and hence will develop a new class of ultrahard and refractory multifunctional materials with advanced electronic, thermal and optical properties for applications at ambient and extreme conditions.

ORCID IDs : Vladimir L. Solozhenko  https://orcid.org/0000-0002-0881-9761